# SECURITY AND ROBUSTNESS ANALYSIS OF ENCRYPTED COMMUNICATIONS FOR LOCAL HOME AUTOMATION: ESPHOME NATIVE API AND MQTT OVER TLS 1.3

Carlos Roberto González-Escarpeta , Juan Eduardo García Sánchez , José Antonio Hernández Reyes and Ricardo Francisco Martínez-González

Tecnológico Nacional de México – IT de Veracruz, Veracruz, Veracruz, México

### ABSTRACT

*The expansion of the Internet of Things (IoT) in residential and local automation environments has highlighted the limitations of considering the internal network as an inherently trusted domain. Man-in-the-middle (MitM) attacks, such as those based on ARP spoofing, can allow traffic interception or manipulation when communications lack appropriate confidentiality, integrity, and authentication mechanisms. This paper presents a systematized review of the security mechanisms used by two communication alternatives between ESP-family microcontrollers and the Home Assistant platform: MQTT protected through TLS 1.3 and the ESPHome Native API, whose encryption mechanism is based on the Noise Protocol Framework using the Noise_NNpsk0_25519_ChaChaPoly_SHA256 pattern. The analysis considers four dimensions: cryptographic robustness, attack surface, implementation complexity, and resource consumption in constrained devices. The review identifies significant differences resulting from the architecture and cryptographic model adopted by each alternative. TLS 1.3 provides a widely standardized framework compatible with public-key infrastructure and formal identity-management mechanisms, although it introduces greater configuration complexity and computational overhead during secure-session establishment. The ESPHome Native API employs a more specialized cryptographic scheme with lower operational complexity, eliminates dynamic algorithm negotiation, and uses ChaCha20-Poly1305 to protect communications. The resource-consumption analysis is based on measurements published by the projects and manufacturers themselves, not on an experimental test bench, and this limitation is explicitly acknowledged. The results identify the conditions under which each mechanism is appropriate for local home-automation architectures and provide a basis for designing IoT systems that require a balance among security, interoperability, and resource constraints.*

## 1. INTRODUCTION

The expansion of the Internet of Things (IoT) in residential, educational, and local automation environments has increased the number of connected devices that continuously exchange information over IP networks. Sensors, actuators, and microcontrollers are part of architectures in which the availability, confidentiality, and integrity of communications are essential to guarantee secure operation. In this context, considering the local network as an inherently trusted environment represents a significant limitation from a security standpoint. Man-in-the-middle (MitM) techniques, such as ARP spoofing, can allow an attacker located within the same network segment to intercept or manipulate the traffic exchanged between automation devices and the server responsible for their control [1], [2]. For this reason, the use of cryptographic mechanisms in internal communications is a relevant element in the design of secure IoT architectures.
Within the Home Assistant ecosystem, microcontrollers of the ESP family can establish communication through different mechanisms that offer distinct cryptographic guarantees and architectural models. Two particularly relevant alternatives are MQTT protected through TLS 1.3 and the ESPHome Native API. Although both allow information exchange to be protected, their communication models, authentication mechanisms, and key management schemes present substantial differences that have not been systematically contrasted under a single set of analysis dimensions. These differences become especially important when end nodes are implemented on resource-constrained microcontrollers, such as the ESP8266 and ESP32, because establishing secure sessions, cryptographic processing, memory usage, and configuration complexity can affect the overall performance of the system.

The objective of this work is to analyze the security and robustness of MQTT over TLS 1.3 and the ESPHome Native API in local home automation architectures, considering four main dimensions: cryptographic robustness, attack surface, implementation complexity, and resource consumption. The purpose is not to establish an absolute superiority between the two alternatives, but to identify their characteristics, limitations, and application scenarios within IoT systems based on Home Assistant and ESP-family microcontrollers. As an application context, this analysis constitutes a technical foundation for the design of a home automation control infrastructure in classrooms at the Tecnológico Nacional de México, Campus Veracruz, where the selection of the communication mechanism must consider both security requirements and the computational constraints of the devices employed.

The specific contributions of this article are the following:

- A systematized review of the cryptographic mechanisms, identity models, and key management procedures employed by MQTT over TLS 1.3 and by the encrypted ESPHome Native API, organized around four explicit research questions and a declared threat model.
- A comparative analysis of both mechanisms across four dimensions —cryptographic robustness, attack surface, implementation complexity, and resource consumption— that distinguishes qualitative properties from quantitative evidence and discusses their comparability.
- A consolidated synthesis of the security advisories and vulnerability records that illustrate how implementation choices, rather than algorithmic choices, determine the practical resistance of cryptographic mechanisms on ESP-family microcontrollers.
- A set of technical criteria for selecting the appropriate communication mechanism in local home automation infrastructures, with particular reference to an educational deployment at the Tecnológico Nacional de México, Campus Veracruz.

The remainder of the article is organized as follows: Section 2 presents the theoretical and technological framework related to MitM attacks, Home Assistant, MQTT, and ESPHome; Section 3 describes the methodology of the systematized review; Sections 4 and 5 describe the cryptographic mechanisms used by MQTT over TLS 1.3 and by the ESPHome Native API, respectively; Section 6 develops the analysis of both alternatives along the established dimensions; Section 7 discusses their implications and application scenarios; and, finally, Section 8 presents the conclusions and future work lines.

# 2. THEORETICAL AND TECHNOLOGICAL FRAMEWORK

## 2.1. Man-in-the-Middle Attacks in IoT Environments

A man-in-the-middle (MitM) attack occurs when an actor manages to position itself logically between two entities that exchange information, with the purpose of observing, intercepting, or modifying communications without the legitimate participants necessarily detecting its presence. In IoT environments, this type of attack becomes especially relevant due to the diversity of devices, protocols, and security mechanisms used within the same infrastructure [1].

In Ethernet or Wi-Fi local networks, a technique that can facilitate this positioning is ARP spoofing. By emitting forged ARP messages, an attacker can associate its own MAC address with the IP address corresponding to another device on the network, such as the router or an automation server. Therefore, part of the traffic destined for the legitimate device can be redirected through the equipment controlled by the attacker [1], [2].

Once positioned in the communication path, the adversary can perform observation and traffic analysis tasks. If the communication lacks adequate cryptographic mechanisms, it can also attempt to modify data or inject messages. The use of encrypted and authenticated channels reduces these risks by providing confidentiality and integrity mechanisms for the transmitted information [2].

Encrypting internal communications is therefore relevant even when devices are located within a local network protected by perimeter mechanisms. For home automation systems, this consideration is particularly important when the exchanged messages can modify the physical state of actuators, lighting systems, access controls, or other automation elements.

## 2.2. Home Assistant Ecosystem

Home Assistant is an automation platform that allows integrating heterogeneous devices, services, and technologies under a primarily local control infrastructure. In this model, the Home Assistant server manages the available entities, receives state changes, executes automations, and transmits commands to the integrated devices.

Peripheral nodes can be implemented using different hardware platforms. Within the scope of this work, ESP8266 and ESP32 microcontrollers are primarily considered, as they are widely used for acquiring variables through sensors and controlling actuators connected to Wi-Fi networks.

These devices present memory, processing capacity, and, in certain scenarios, energy consumption constraints. Such limitations make it necessary to consider not only the cryptographic robustness of the mechanisms used, but also the number of resources required to establish and maintain secure communications.

Home Assistant can interact with these devices through different mechanisms. Among them are MQTT, based on a publish-subscribe model through an intermediary, and the ESPHome Native API, designed to establish a direct integration between devices running ESPHome firmware and Home Assistant.

## 2.3. MQTT: Publish-Subscribe Architecture

MQTT (Message Queuing Telemetry Transport) is a messaging protocol based on the publish-subscribe model. Its architecture uses an intermediary server called a broker, to which clients connect to publish information or subscribe to specific topics [3].

Publishers post messages associated with a topic, while the broker determines which clients are subscribed and distributes the corresponding messages. This separation between producers and consumers allows decoupling the components of an IoT architecture and facilitates multiple systems receiving the same information without the source device needing to maintain an independent connection with each consumer.

MQTT provides different Quality of Service (QoS) levels. QoS 0 provides delivery under the "at most once" criterion; QoS 1 seeks to deliver the message "at least once"; and QoS 2 incorporates an additional exchange intended to provide "exactly once" delivery. The protocol also has mechanisms such as retained messages and persistent sessions that facilitate different information distribution strategies [3].

It is important to note that MQTT does not determine the format of the transported content. The payload of a PUBLISH message corresponds to application-defined data and can contain textual or binary information. Therefore, formats such as JSON can be used in certain implementations, but they are not a requirement of the MQTT protocol.

From a cryptographic standpoint, MQTT does not provide payload encryption by itself. Communication protection can be established through mechanisms provided by the transport layer, with TLS being one of the alternatives used to provide confidentiality, integrity, and authentication of the channel between the client and the broker [4]. This work specifically analyzes MQTT protected through TLS 1.3.

## 2.4. ESPHome Native API: Direct Communication with Home Assistant

ESPHome provides a native API designed specifically for communication between devices running ESPHome firmware and Home Assistant. Unlike the MQTT architecture, this mechanism does not require an intermediary broker for ordinary information exchange between the two endpoints.

In this architecture, the ESPHome device exposes a TCP service, and Home Assistant establishes and maintains a persistent connection with the node. The default port used by the Native API is 6053 [5]. This bidirectional connection allows the device to report state changes as they occur and allows Home Assistant to transmit commands to its entities.

The API uses Protocol Buffers for message serialization [5]. This mechanism provides a structured binary representation and employs, among other resources, variable-length integer encoding (VarInt), in which small values can be represented using a reduced number of bytes.

The Native API specification contemplates both unencrypted communication and communication protected through the Noise Protocol Framework. For this reason, the use of the Native API should not be considered equivalent, by itself, to encrypted communication. For the purposes of this work, only its configuration with Noise encryption enabled through a pre-shared key is analyzed.

This distinction is fundamental for the subsequent analysis. MQTT and the ESPHome Native API respond to different architectural models: MQTT provides decoupling between producers and consumers through a broker, while the Native API favors a direct and specific integration with Home Assistant. Consequently, the differences in security, resource consumption, and implementation complexity do not depend solely on the cryptographic algorithm used, but also on the communication architecture on which each mechanism operates. Figure 1 illustrates both communication architectures side by side, highlighting the presence of the broker in the MQTT model and the direct point-to-point connection in the ESPHome Native API model.

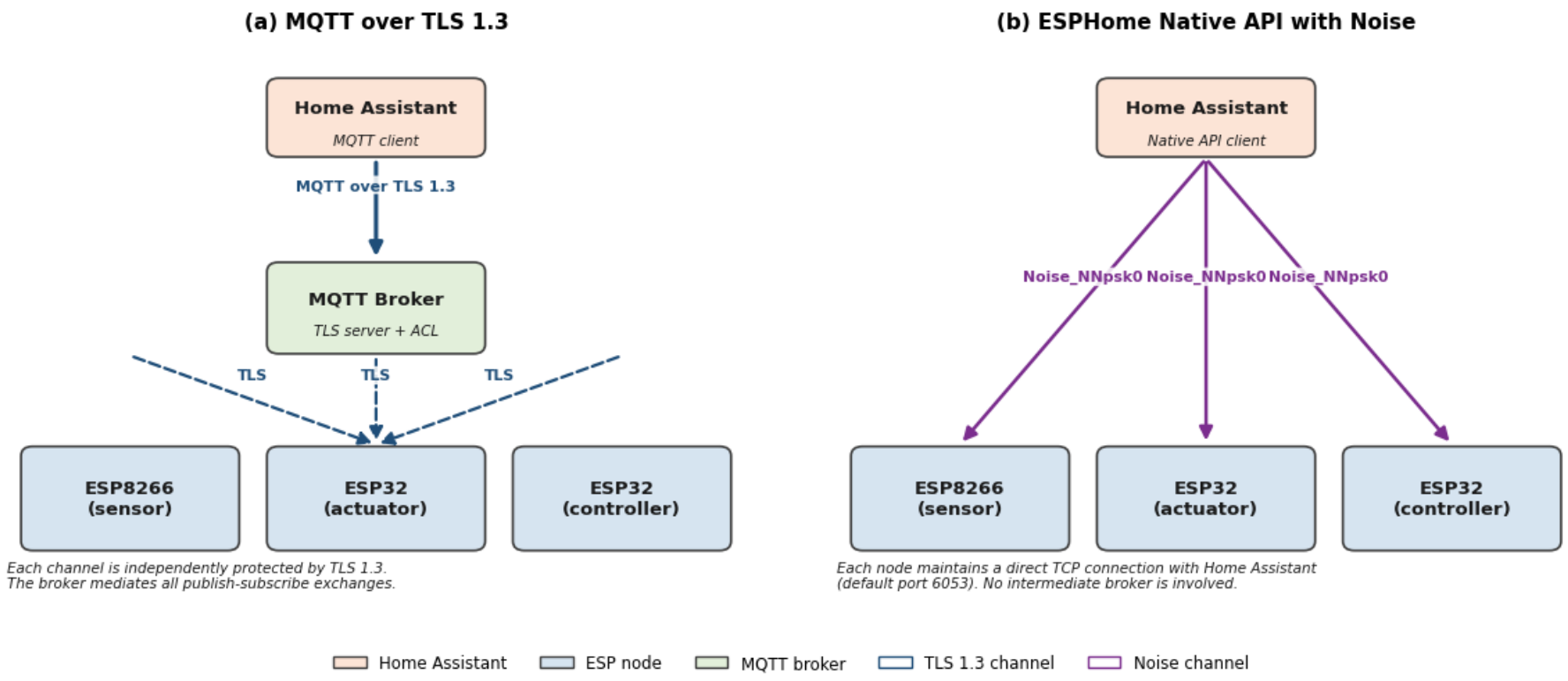

Figure 1. Communication architectures of the two mechanisms analyzed. (a) MQTT over TLS 1.3: ESP nodes connect to a central broker that mediates publish-subscribe exchanges and concentrates authentication, authorization, and message distribution; TLS protects each client-to-broker channel. (b) ESPHome Native API with Noise encryption: Home Assistant establishes a direct persistent TCP connection (default port 6053) with each ESP node, without an intermediate broker; Noise_NNpsk0 protects the point-to-point channel. Both diagrams assume operation within the same local network segment, consistent with the threat model declared in Section 3.5.

## 3. METHODOLOGY

This work constitutes a systematized review [6], [7] whose purpose is to compare two encrypted communication mechanisms used in local home automation architectures based on ESP-family microcontrollers and Home Assistant: MQTT over TLS 1.3 and the ESPHome Native API with Noise-based encryption. A systematized review adopts an explicit and reproducible search and selection procedure but does not claim the exhaustiveness of a full systematic review. This approach is appropriate when the object of study is a rapidly evolving technical domain in which peer-reviewed literature is still limited and a significant part of the relevant evidence is found in standards, protocol specifications, and official technical documentation [8].

### 3.1. Research Questions

The review is guided by four research questions, aligned with the four analysis dimensions declared in Section 1:

- RQ1. What cryptographic mechanisms, key establishment procedures, and identity models are employed by MQTT over TLS 1.3 and by the encrypted ESPHome Native API?
- RQ2. What security properties (confidentiality, integrity, authentication, forward secrecy, and post-quantum readiness) does each mechanism provide, and under what threat model?

- RQ3. What resource costs (memory footprint, handshake time, per-message overhead, and computational load) have been reported for each mechanism on ESP8266 and ESP32 platforms?
- RQ4. In what application scenarios within a local home automation infrastructure is each mechanism appropriate?

Sections 4 and 5 address RQ1 and RQ2, Section 6 addresses RQ3, and Sections 7 and 8 address RQ4.

### 3.2. Sources and Search Strategy

The search covered four types of sources, selected according to the nature of the evidence available for each mechanism:

- Peer-reviewed scientific literature, consulted through IEEE Xplore, ACM Digital Library, Scopus, Web of Science, and arXiv. The search strings combined terms related to the platforms (ESP8266, ESP32, ESPHome, Home Assistant), the protocols (MQTT, TLS 1.3, Noise Protocol), and the analysis dimensions (security, performance, resource consumption, embedded).
- Standards and protocol specifications, including RFC 9846 (TLS 1.3) [11], RFC 10024 (post-quantum hybrid key agreement for TLS 1.3) [19], the Noise Protocol Framework specification [12], and the OASIS MQTT 5.0 standard [3].
- Official technical documentation from the projects and manufacturers involved: ESPHome API protocol details [5], the ESPHome 2026.9.0 release announcement [20], and the ESP-IDF Mbed TLS programming guide [21].
- Security advisories and vulnerability records, including CVE-2024-45191 [15], CVE-2018-0498 [16], and the Espressif advisories AR2022-003 [17] and AR2026-005 [18].

The inclusion of technical documentation and advisories as primary sources is justified because the ESPHome Native API with Noise encryption has not yet been the subject of extensive peer-reviewed evaluation, and because the resource measurements published by the projects themselves constitute the only currently available quantitative evidence for that mechanism on ESP platforms. This decision is consistent with the systematized review approach, which permits the incorporation of gray literature when it is the most reliable available source, provided that its nature is declared and its limitations are discussed.

### 3.3. Inclusion and Exclusion Criteria

Sources were included if they met all the following criteria:

- They addressed at least one of the two mechanisms under study, or a cryptographic primitive directly relevant to them (X25519, ChaCha20-Poly1305, AES-GCM, SHA-256).
- They were published between 2018 and 2026, a period that covers the standardization of TLS 1.3, the consolidation of the Noise Protocol Framework, and the introduction of post-quantum hybrid key agreement for TLS 1.3.
- They provided technical specifications, security properties, or quantitative measurements relevant to at least one of the four research questions.
- They were written in English or Spanish.

Sources were excluded if they:

- Addressed generic IoT security without reference to MQTT, TLS, Noise, or the ESP platform.

- Reported performance figures without specifying hardware, firmware version, or cryptographic configuration.
- Were commercial product descriptions or marketing material without technical content.
- Duplicated information already covered by a primary source (for example, secondary summaries of RFC 9846).

### 3.4. Data Extraction and Synthesis

For each included source, the following information was extracted: type of source, mechanism addressed, cryptographic primitives involved, security properties reported, hardware and software context (when applicable), metrics reported, and limitations declared by the authors.

The synthesis was organized around the four research questions and the four analysis dimensions. Because the available quantitative evidence for the two mechanisms comes from different platforms and configurations, the review does not perform a direct numerical comparison. Instead, it presents the values in their original context and discusses their comparability explicitly in Section 6.4. This decision responds to the methodological caution expressed by Restuccia et al. [9] and Prantl et al. [10] regarding the risk of comparing TLS performance figures obtained under heterogeneous conditions.

### 3.5. Threat Model Considered

The security analysis assumes a threat model consistent with the application context described in Section 1: an adversary with logical access to the same local network segment as the devices, capable of performing passive traffic observation and active man-in-the-middle attacks based on ARP spoofing. The adversary is assumed to be unable to obtain physical access to the devices or to compromise the Home Assistant server. Physical attacks, supply-chain attacks, and adversaries with quantum computing capabilities are out of scope, although the post-quantum readiness of each mechanism is discussed qualitatively as part of RQ2, following the recent standardization of hybrid key agreement for TLS 1.3 [19].

### 3.6. Limitations of the Review

Three limitations follow from the methodology described above. First, the evidence for the ESPHome Native API relies substantially on official project documentation and release announcements rather than on independent peer-reviewed studies; consequently, the reported resource figures reflect the measurements published by the project itself and have not been independently replicated. Second, the quantitative values for MQTT over TLS 1.3 and for the ESPHome Native API come from different hardware platforms, firmware versions, and library configurations, so they are not directly comparable; this review therefore presents them in context and does not derive a numerical ranking from them. Third, the review does not include an experimental validation on a common test bench, which is identified in Section 8 as the principal line of future work.

## 4. CRYPTOGRAPHIC PROTECTION OF MQTT THROUGH TLS 1.3

As noted above, MQTT does not incorporate its own cryptographic mechanisms to protect message confidentiality. When a secure communication between an MQTT client and the broker is required, a widely used alternative is to run MQTT over Transport Layer Security (TLS). In this configuration, TLS establishes a protected channel before the ordinary exchange of MQTT messages, providing confidentiality and integrity of data in transit and mechanisms to authenticate the communication endpoints.

This work specifically considers TLS 1.3, currently specified in RFC 9846 [11]. This version preserves the design originally introduced by TLS 1.3 and restricts record protection to Authenticated Encryption with Associated Data (AEAD) algorithms, eliminating various mechanisms inherited from previous versions.

## 4.1. Secure Channel Establishment through TLS 1.3

Before the MQTT client and the broker exchange the messages corresponding to the application protocol, TLS establishes the necessary cryptographic parameters through a negotiation process called the handshake.

In a full TLS 1.3 session establishment, the client initially transmits a ClientHello message containing, among other parameters, the supported cryptographic options and the information necessary for key exchange. The server responds with a ServerHello message selecting the session parameters. From this point on, the subsequent messages of the establishment process are protected using the keys derived during the exchange [11].

TLS 1.3 simplified the procedure compared to TLS 1.2 and allows completing an ordinary session establishment in one round trip (1-RTT) when a HelloRetryRequest is not necessary [11]. It also contemplates session resumption mechanisms through pre-shared keys (PSK) and an early data transmission mode called 0-RTT. The latter reduces reconnection latency, although it presents different security properties, particularly regarding the possibility of message replay, so its use must be evaluated according to the characteristics of the application [11].

Application traffic protection in TLS 1.3 is performed exclusively through AEAD algorithms. Among the defined suites are AES-GCM and ChaCha20-Poly1305 [11]. This scheme allows transmitted information to be encrypted and authenticated, so that unauthorized modification of protected records can be detected by the receiver.

Another relevant characteristic is the use of ephemeral key exchange mechanisms when public-key cryptography is employed. These mechanisms allow providing forward secrecy, so that the eventual later exposure of a long-term authentication key does not necessarily imply the recovery of the session keys previously used [11].

## 4.2. Broker and Client Authentication

The protection provided by TLS must be distinguished from MQTT's own authorization mechanisms. TLS establishes and protects the communication channel, while MQTT can apply additional mechanisms to determine whether a client is authorized to publish or subscribe to specific topics [3].

A common deployment uses server authentication through digital certificates. In this scheme, the broker presents its certificate during the TLS establishment, and the client verifies its validity and the server's identity before continuing with the connection. X.509 certificates constitute the most common mechanism for this purpose, although TLS 1.3 also contemplates other key establishment models, including pre-shared keys [11].

Once the TLS channel is established, MQTT can use username and password credentials in the CONNECT message. Since the message is encapsulated within the protected channel, these credentials are not transmitted in plaintext over the network. However, their security depends on proper password management and the implementation of authorization mechanisms in the broker, such as access control lists (ACLs) associated with topics [3], [4].
An alternative providing stronger device identity binding consists of mutual authentication through certificates (mTLS). In this configuration, both the broker and the client present cryptographic credentials during the TLS process. This mechanism allows associating the device's identity directly with its certificate, although it increases operational complexity due to the processes necessary for certificate provisioning, storage, renewal, and revocation [4].

Therefore, the use of TLS does not by itself determine a single authentication model for MQTT. The architecture can combine server authentication through certificates, application-level credentials, client certificates, or PSK-based mechanisms, depending on the system requirements.

# 5. Cryptographic Protection of the ESPHome Native API through Noise

The ESPHome Native API can operate through unencrypted communication or through a protected channel based on the Noise Protocol Framework. Within the scope of this work, only the configuration with encryption enabled is considered, since the objective is to analyze communication mechanisms capable of providing confidentiality and integrity against an adversary with the ability to observe or manipulate local network traffic.

Unlike MQTT, whose cryptographic protection is established through an additional transport layer such as TLS, ESPHome integrates the Noise mechanism within the protocol used by its Native API. The current implementation employs the Noise_NNpsk0_25519_ChaChaPoly_SHA256 pattern, which combines a pre-shared key with an ephemeral key exchange and authenticated encryption [5].

## 5.1. Noise Protocol Framework and the NNpsk0 Pattern

The Noise Protocol Framework is a framework for constructing interactive cryptographic protocols from previously defined patterns [12]. Each pattern specifies the sequence of operations necessary to establish session keys and determines the authentication and confidentiality properties obtained during the exchange. In the case of ESPHome, the Noise_NNpsk0_25519_ChaChaPoly_SHA256 pattern is composed of the following elements:

- NN: indicates an establishment pattern in which neither participant uses a static public key as an identity mechanism during the exchange. Both endpoints generate ephemeral keys to establish the session secret [12].
- psk0: incorporates a pre-shared key (PSK) at the beginning of the first message of the protocol. In ESPHome, this key contains 32 bytes and is normally provided encoded in Base64 through the encryption.key parameter of the API configuration [5].
- 25519: identifies the use of X25519, based on Curve25519, to perform the Diffie-Hellman exchange between the ephemeral keys generated by both participants [12].
- ChaChaPoly: specifies ChaCha20-Poly1305 as the AEAD algorithm to protect the data exchanged after session establishment [12].
- SHA256: identifies SHA-256 as the hash function used by the Noise symmetric state and by the key derivation mechanisms associated with the protocol [12].

The combination of the PSK with ephemeral material derived through X25519 prevents the stored key from being used directly as the encryption key for messages. Transport keys are derived during each session establishment. The use of ephemeral keys introduces forward secrecy properties;

however, these properties must be expressed in terms of the threat model and the moment at which PSK exposure could occur. In particular, the NNpsk0 pattern does not unconditionally provide the same identity, or forward secrecy guarantees as other Noise patterns that use authenticated static keys [12].

This difference is relevant compared to TLS with certificates. In the pattern used by ESPHome, there is no PKI infrastructure or X.509 certificate chain. Trust between Home Assistant and the device is established primarily through knowledge of the shared PSK [5].

## 5.2. ChaCha20-Poly1305 and Its Relationship with AES

AES and ChaCha20 are widely used cryptographic primitives, although they respond to different design structures. AES is a standardized block cipher extensively used in general-purpose systems, communication protocols, and hardware cryptographic accelerators [13]. ChaCha20, on the other hand, is a stream cipher built primarily through addition, rotation, and XOR (ARX) operations [14]. This difference should not be interpreted as an inherent cryptographic weakness of AES. Its security depends, among other factors, on the implementation employed. Some software implementations of AES based on lookup tables can generate memory access patterns dependent on secret information. When such accesses produce observable differences in cache memory behavior, timing side channels can originate [15].

A documented example is CVE-2024-45191, corresponding to the AES implementation of the libolm library. In this case, the use of lookup tables for operations related to the S-boxes allowed cache-based timing attacks. The vulnerability record includes references to a public technical demonstration. This case does not imply a vulnerability of the AES algorithm, but rather of a specific implementation that did not guarantee behavior independent of secret data [15].

There are also precedents related to embedded platforms. CVE-2018-0498 affected versions of Mbed TLS through a side channel associated with cache behavior in certain CBC suites. Although this vulnerability does not correspond to TLS 1.3 nor demonstrates a general weakness of AES-GCM, it illustrates the importance of the concrete implementation of cryptographic primitives in resource-constrained devices [11].

Hardware implementations also do not automatically eliminate side-channel risks. Espressif documented in advisory AR2022-003 physical side-channel attacks capable of recovering cryptographic material associated with the AES core of certain revisions of the ESP32 family [17]. More recently, advisory AR2026-005 documented AES-128 key recovery from the AES peripheral of the ESP32-S3 through voltage fault injection and differential fault analysis [13].

These attacks require physical access, instrumentation, and capabilities considerably different from those necessary for a network MitM attack. Nevertheless, they constitute evidence that the resistance of a cryptographic system depends both on the algorithm and on its software or hardware implementation and on the threat model considered.

ChaCha20 presents a favorable characteristic for software implementations on processors that lack specific cryptographic acceleration. Its ARX construction avoids the use of S-boxes and lookup tables indexed by secret information, facilitating implementations with execution times less dependent on the processed data [14].

ESPHome uses ChaCha20 together with Poly1305. ChaCha20 provides confidentiality, while Poly1305 generates a cryptographic tag that allows verifying the authenticity and integrity of the message. The ChaCha20-Poly1305 combination constitutes an AEAD scheme, so an unauthorized modification of the ciphertext causes verification failure before the message is accepted by the application [5], [12].

Therefore, the choice of ChaCha20-Poly1305 in the ESPHome Native API is particularly suitable for an implementation intended for different generations of microcontrollers, including those that do not have AES accelerators. This consideration does not by itself allow establishing a general superiority of ChaCha20 over AES, since relative performance depends on the processor architecture and the availability of hardware cryptographic acceleration [14].

## 5.3. Structure of Encrypted Frames and Message Protection

Once the Noise establishment is completed, the ESPHome Native API encapsulates Protocol Buffers messages within frames encrypted using ChaCha20-Poly1305.

The transport frame contains a one-byte indicator with the value 0x01, followed by a two-byte field that specifies the size of the encrypted information. Subsequently, the message type, its length, and the data serialized through Protocol Buffers are found, all of them cryptographically protected. Finally, a 16-byte authentication tag generated through Poly1305 is added [5].

ESPHome documentation describes a structure in which the control and authentication fields represent a fixed overhead of 23 bytes per message: seven bytes correspond to the logical header considered by the format, and sixteen to the authentication tag [5].

It should be noted that the two-byte field expressing the size of the encrypted payload is transmitted unencrypted. Consequently, Noise protects the content, the message type, and the internal length of the data, but a network observer can still obtain certain traffic information, such as the moment when frames are exchanged and their approximate size. Encryption provides content confidentiality, but it does not intend to completely eliminate traffic metadata analysis.

The Poly1305 tag allows detecting modifications to the encrypted information. If authentication of a frame fails, the message cannot be accepted as valid, and the implementation terminates the connection. In this way, an attacker positioned in the middle of the communication can observe or block packets but should not be able to transparently modify the protected content without causing an authentication failure [5].

## 5.4. Key Management and Limitations of the Security Model

The operational simplicity of the ESPHome encrypted API comes partly from replacing the certificate infrastructure with a 32-byte PSK shared between the device and authorized clients. This characteristic reduces provisioning complexity in local installations but introduces specific key management requirements.

The security of the mechanism depends on maintaining the confidentiality of the PSK. Each device should preferably have a different key, avoiding reusing the same secret across multiple nodes. ESPHome documentation recommends using unique keys and storing them outside configuration files that could be incorporated into public repositories [5].

Exposure of the device configuration, of files containing secrets, or of the non-volatile memory where the PSK is stored can compromise the authentication mechanism corresponding to that node.

On compatible devices, additional storage protection mechanisms can reduce the risk associated with physical credential extraction.
On the other hand, the NNpsk0 pattern does not provide an identity based on certificates or static public keys verified through an external infrastructure. The practical identity of the device is linked to knowledge of the PSK and to the process by which said key was provisioned [5], [12].

These characteristics make the scheme especially appropriate for locally managed architectures in which Home Assistant and ESPHome nodes are under the same management domain. However, its trust model should not be considered equivalent to that provided by a PKI infrastructure oriented toward organizations with formal identity management, certification, and revocation requirements. Consequently, the security analysis of the Native API cannot be limited to comparing ChaCha20-Poly1305 with the algorithms used by TLS. The Noise pattern employed, PSK management, physical protection of the device, communication architecture, and the capabilities of the hardware on which it runs must be considered jointly.

## 6. SECURITY, ARCHITECTURE AND EFFICIENCY ANALYSIS

MQTT protected through TLS 1.3 and the encrypted ESPHome Native API provide mechanisms capable of protecting the confidentiality and integrity of communications; however, they respond to different architectural and trust management models. For this reason, their analysis should not be reduced exclusively to the cryptographic algorithms employed.

This section considers the four dimensions established as the scope of the work: cryptographic robustness, attack surface, implementation complexity, and resource consumption. Table 1 consolidates the qualitative comparison across all four dimensions and provides a single reference for the discussion that follows.

Table 1. Comparative summary of MQTT over TLS 1.3 and the ESPHome Native API across the four analysis dimensions.

| Dimension | MQTT over TLS 1.3 | ESPHome Native API (Noise) |
|---|---|---|
| Cryptographic primitives | AES-GCM or ChaCha20-Poly1305 (negotiated) | ChaCha20-Poly1305 (fixed) |
| Key establishment | ECDHE (X25519, P-256), PSK, hybrid PQC groups | X25519 ephemeral + PSK of 32 bytes |
| Identity model | X.509 certificates, mTLS, or PSK | Symmetric PSK, no PKI |
| Forward secrecy | Yes, with ephemeral ECDHE | Partial; depends on threat model and PSK exposure timing |
| Post-quantum readiness | Yes; RFC 10024 hybrid groups available in TLS 1.3 | No equivalent hybrid mechanism in the current pattern |
| Trust anchor | PKI or pre-shared secret | Pre-shared secret |
| Communication model | Publish-subscribe with broker | Direct point-to-point with Home Assistant |
| Attack surface | Broker and clients; compromise of broker affects multiple devices | Per-device PSK; compromise of one node limited if keys are unique |
| Implementation complexity | High: broker, certificates, ACLs, TLS parameters | Low: single 32-byte key and predetermined suite |
| Key management | Formal lifecycle (issuance, renewal, revocation) | Manual provisioning and rotation of PSKs |

| Dimension | MQTT over TLS 1.3 | ESPHome Native API (Noise) |
|---|---|---|
| Handshake cost (ESP8266) | Not reported in the reviewed literature for TLS 1.3 | ~300 ms (ESPHome 2026.9.0) |
| Handshake cost (ESP32) | ~42 KB heap (Mbed TLS, default configuration) | ~46 ms (ESPHome 2026.9.0) |
| Flash footprint | Not reported in the reviewed literature | 41.9 KB (ESP8266), 37.7 KB (ESP32) with encryption enabled |
| Per-message overhead | Depends on TLS record and MQTT header | 23 bytes fixed (7 header + 16 Poly1305 tag) |
| Interoperability | High; suited for multiple producers and consumers | Specific to Home Assistant |
| Suitable scenario | Multiple consumers, cross-platform integration | Direct node-to-Home Assistant integration |

## 6.1. Cryptographic Robustness

TLS 1.3 constitutes a general-purpose security protocol with cryptographic negotiation capability, different authentication mechanisms, and an ecosystem oriented toward multiple application scenarios [11]. The encrypted ESPHome API uses, in contrast, a specific and currently fixed cryptographic configuration: Noise_NNpsk0_25519_ChaChaPoly_SHA256, formed by X25519, ChaCha20-Poly1305, SHA-256, and a 32-byte pre-shared key [5].

The qualitative synthesis can be expressed in five main differences. In cryptographic architecture, TLS 1.3 supports negotiation within the mechanisms permitted by the version, while ESPHome employs a fixed NNpsk0/25519/ChaChaPoly/SHA256 configuration. In identity, TLS can use X.509 certificates, mTLS, or PSK depending on the implementation [11], while ESPHome bases trust on a shared PSK without PKI [5]. TLS offers general-purpose interoperability, and ESPHome is oriented specifically toward integration with Home Assistant. TLS already has hybrid post-quantum transition mechanisms standardized for TLS 1.3 by RFC 10024 [19], although this does not necessarily imply that these mechanisms are available in the Mbed TLS implementations used on the analyzed ESP devices. Finally, TLS can be integrated with formal certificate management processes, while ESPHome privileges direct management of pre-shared keys.

A particularly relevant aspect for cryptographic evolution is the post-quantum transition. In August 2026, RFC 10024 standardized for TLS 1.3 the hybrid groups X25519MLKEM768, SecP256r1MLKEM768, and SecP384r1MLKEM1024, which combine a traditional ECDHE exchange with ML-KEM [19]. The current ESPHome implementation continues to use X25519, ChaChaPoly, and SHA-256 without incorporating an equivalent hybrid post-quantum mechanism. This asymmetry does not imply an immediate vulnerability of the ESPHome scheme, since no quantum adversary is included in the threat model declared in Section 3.5, but it does represent a difference in the capacity for future cryptographic evolution that should be considered in long-lived deployments.

This difference does not imply that one of the alternatives is intrinsically secure and the other insecure. TLS prioritizes flexibility, interoperability, and cryptographic evolution, while ESPHome deliberately reduces the number of cryptographic options to simplify implementation and reduce the possibility of errors derived from algorithm selection.

## 6.2. Attack Surface

Architectural differences also modify the attack surface. In MQTT, devices establish their connections with a broker, which concentrates message distribution between producers and consumers. The existence of this central component provides administrative advantages, since it allows concentrating authentication, authorization, logs, and access policies. At the same time, an incorrect configuration or compromise of the broker can have consequences for multiple connected devices [3], [4].

The MQTT specification recommends applying authorization controls that limit the topics on which each client can publish or subscribe, paying special attention to broad-scope filters such as # [3]. Therefore, the possibility that a compromised client accesses information from other devices is not an inevitable consequence of MQTT but depends largely on the ACLs and the authorization model configured in the broker [4].

In the ESPHome Native API, Home Assistant establishes connections directly with each node. The absence of the broker eliminates an intermediary component from the ordinary communication flow and allows credentials and the state of a connection to be associated with a specific device. This model can limit the scope of exposure derived from the compromise of a node's PSK when each device has a different key. However, the architecture continues to depend on Home Assistant as the orchestrating platform, so compromise of the central server still represents a high-impact scenario [4].

Similarly, the security provided by compartmentalization decreases if the same PSK is reused across multiple devices. Therefore, the benefit of the point-to-point model also depends on adequate individual key management.

In both cases, TLS and Noise protect the content of communications, but they do not eliminate attacks oriented toward availability. An adversary with the ability to block, delay, or discard traffic can cause interruptions even if it cannot validly decrypt or modify messages [2].

### 6.3. Implementation and Management Complexity

An MQTT infrastructure protected through TLS may require broker configuration, certificates or keys, authentication mechanisms, ACL policies, and TLS parameters. This complexity allows building architectures with multiple producers and consumers, segmenting privileges, and using formal identity management mechanisms [4].

Therefore, this complexity should not be considered exclusively as a disadvantage: part of it responds to interoperability and administration functionalities that are not part of the design objective of the ESPHome Native API.

In ESPHome, enabling API encryption essentially requires providing an encryption key and sharing it with Home Assistant. The protocol uses a 32-byte PSK, and the cryptographic suite is predetermined, reducing the number of cryptographic decisions the administrator must make [5]. This simplicity is favorable in local installations with a limited number of devices managed under the same domain. However, as the number of nodes increases, so does the need to maintain key inventories, perform rotations, and replace secrets in case of compromise.

The difference can therefore be expressed as a trade-off between provisioning simplicity and a formal identity management infrastructure. ESPHome favors the former aspect, while MQTT over TLS can be integrated with mechanisms of greater scale and administrative granularity.

## 6.4. Resource Consumption and Efficiency

Quantitative comparison of computational cost requires special caution. The values published for TLS and Noise frequently come from different platforms, libraries, cryptographic configurations, and firmware versions. Consequently, it is not methodologically correct to present those results as if they had been obtained through the same controlled experiment [9], [10].

In particular, the throughput figures previously used that could not be verifiably attributed to the microcontrollers considered are removed from this analysis. Instead, specific measurements from ESPHome and ESP-IDF implementations are presented, preserving their context of origin. Figure 2 consolidates the three-resource metrics reported by ESPHome for the ESP8266 and ESP32 platforms, showing the evolution between versions 2026.8.0 and 2026.9.0.

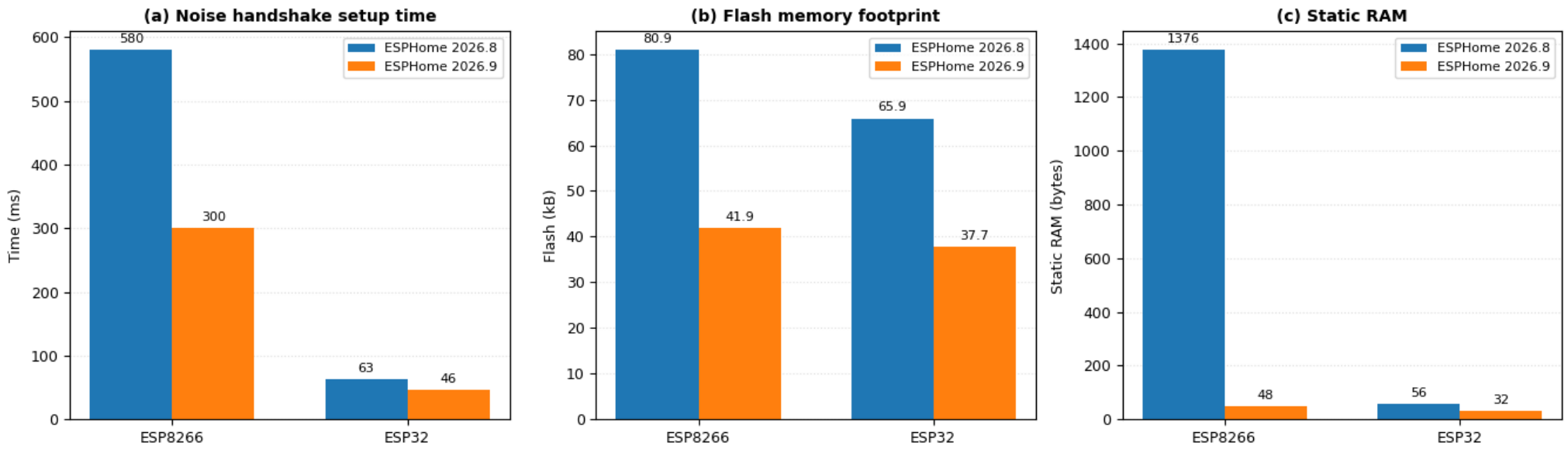

Figure 2. Resource metrics associated with API encryption in ESPHome: (a) Noise handshake establishment time (ms), (b) Flash memory footprint (kB), and (c) static RAM (bytes). Data compare versions 2026.8.0 and 2026.9.0 on ESP8266 d1 mini and ESP32 AtomU platforms, as published in the ESPHome 2026.9.0 release announcement [20]. The three panel report the same release comparison and should be interpreted together; the values correspond to the increment attributable to API encryption and not to per-connection dynamic memory.

ESPHome 2026.9.0 reports that the Noise handshake establishment time decreased approximately from 580 ms to 300 ms on ESP8266 and from 63 ms to 46 ms on ESP32. In the same release, the Flash memory cost associated with API encryption decreased from 80.9 kB to 41.9 kB on ESP8266 d1 mini and from 65.9 kB to 37.7 kB on ESP32 AtomU, while static RAM decreased from 1376 bytes to 48 bytes on ESP8266 d1 mini and from 56 bytes to 32 bytes on ESP32 AtomU [20]. These figures show that the platform and the implementation version directly influence the session establishment cost and the memory footprint, and that the differences between platforms are large enough that comparisons should be made within each platform rather than across platforms.

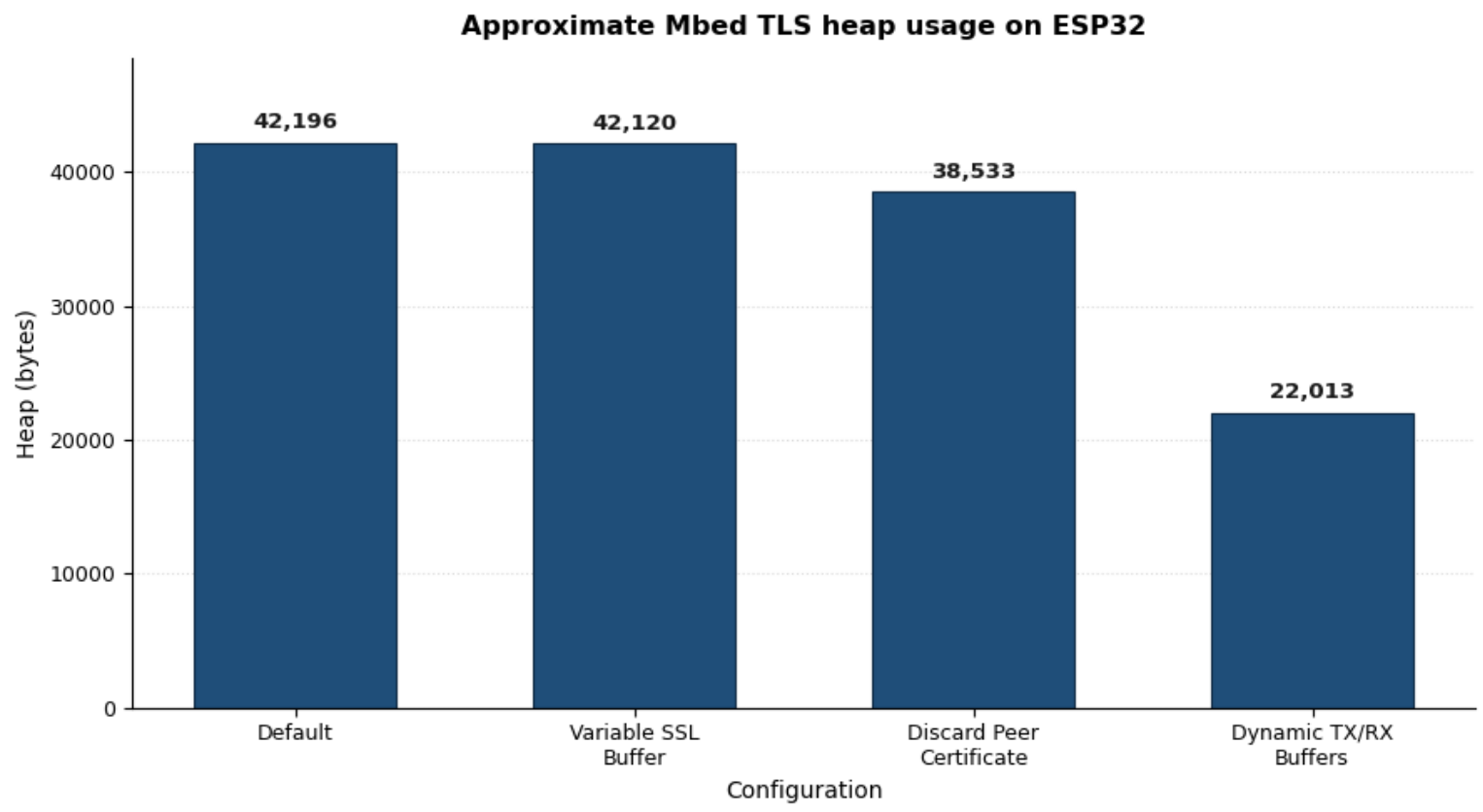

Figure 3. Approximate Mbed TLS heap usage on ESP32 under four configurations. Data from the ESP-DF Mbed TLS programming guide [21]. Measurements correspond to a test with Mbed TLS and server validation on ESP32 and depend on the library version and the configuration parameters; they should not be interpreted as a generic representation of the cost of TLS 1.3 on all ESP32 variants.

The documentation of ESP-IDF illustrates the variability of TLS cost. In a test with Mbed TLS and server validation, Espressif reports approximately 42,196 bytes of heap with the default configuration, 42,120 bytes with variable SSL buffer length, 38,533 bytes when not retaining the peer certificate, and 22,013 bytes when enabling dynamic TX/RX buffers. The documentation itself warns that these values change according to the configuration and the Mbed TLS version [21].

On the other hand, the Noise transport format used by ESPHome introduces a fixed control and authentication structure in each message. This characteristic makes it easy to estimate its protocol overhead, although it does not replace a comprehensive traffic measurement that includes TCP, Wi-Fi, and application content.

These data suggest that the encrypted ESPHome API has been the subject of specific optimizations for ESP microcontrollers, particularly relevant on the ESP8266. However, they do not by themselves allow asserting that Noise is quantitatively more efficient than TLS 1.3 in all scenarios, because they do not come from a common test bench.

To establish a rigorous quantitative comparison, it would be necessary to run both mechanisms on the same hardware, using the same network, payload size, transmission frequency, and connection conditions. At least the following should be measured: establishment time, free memory before and after connection, processing time per message, total volume of transmitted data, and energy consumption.

For this reason, the results of this section should be interpreted as a technical analysis supported by specifications and published measurements, while direct experimental comparison between both mechanisms is proposed as a subsequent stage of the research.

## 7. Discussion

The analysis performed shows that MQTT protected through TLS 1.3 and the encrypted ESPHome Native API respond to different architectural objectives, even though both mechanisms can be used to protect communications within a local home automation infrastructure. The observed differences are not reduced to the cryptographic algorithm employed, but involve the communication model, identity management, key administration, interoperability, and the constraints of the end devices.

Consequently, the selection of one mechanism or the other must be made considering the characteristics of the complete system and not solely isolated performance or cryptographic robustness parameters.

## 7.1. Application Scenarios

The encrypted ESPHome Native API presents characteristics particularly suitable for architectures in which Home Assistant constitutes the main orchestrator and ESPHome devices are managed within the same trust domain.

In this scenario, direct communication between Home Assistant and each node eliminates the need to deploy and administer an additional broker. The use of an individual PSK and a predetermined cryptographic suite also reduces the number of parameters that must be configured during device provisioning [5].

The optimizations recently incorporated into ESPHome reinforce this approach on resource-constrained platforms. In ESPHome 2026.9, the reported time for Noise establishment decreased approximately from 580 to 300 ms on ESP8266 and from 63 to 46 ms on ESP32, in parallel with considerable reductions in Flash memory and static RAM associated with the cryptographic implementation [20].
These characteristics are relevant in installations consisting mainly of ESP sensors and actuators whose function is to exchange states and commands directly with Home Assistant.

MQTT over TLS 1.3, on the other hand, offers advantages when the architecture requires decoupling producers and consumers. A single device can publish information that is subsequently processed by Home Assistant, storage systems, analytics platforms, external applications, or other consumers without the need to establish a direct connection with each of them [3].

The existence of a broker introduces an additional component that must be protected and administered, but it allows centralizing authentication, authorization, and message distribution controls. In architectures with multiple consumers, cross-platform integration, or information exchange between different subsystems, these capabilities can justify the additional complexity of the model [4].

Therefore, the fundamental difference should not be posed as an absolute superiority of one mechanism over the other. The Native API favors a direct and specific integration with Home Assistant, while MQTT prioritizes decoupling and interoperability between heterogeneous components.

## 7.2. Local Network Security and Trust Model

The results support the need to cryptographically protect communications even when devices operate within a local network.

The presence of a perimeter firewall does not eliminate the possibility that an adversary gains access to the internal segment through a compromised device, an incorrect configuration, or network-level spoofing techniques [1]. Under these conditions, the authenticated encryption mechanisms provided by TLS 1.3 and Noise reduce the possibility that an attacker can validly interpret or modify exchanged messages [2].

However, channel encryption does not eliminate all risks. Both TLS and Noise still allow an adversary to observe certain metadata, such as transmission timing, flow direction, or approximate size. In the Noise protocol used by ESPHome, for example, the encrypted payload size is part of the unencrypted frame header [5].

Nor does cryptography prevent a denial-of-service attack based on blocking, delaying, or discarding packets. Therefore, cryptographic protection should be considered as one of the components of a broader security architecture that includes network segmentation, access control, firmware updates, and protection of credentials stored on devices.

## 7.3. Identity and Key Management

One of the most relevant contrasts between the two mechanisms lies in trust management.
TLS allows the use of X.509 certificates and mutual authentication, which facilitates integration with formal identity infrastructures and credential issuance, renewal, and revocation processes [11]. This capability increases administrative complexity, but it may be necessary in larger-scale installations or those subject to organizational control and audit requirements [4].

ESPHome simplifies this process by using a 32-byte PSK. This model significantly reduces provisioning effort but shifts part of the security to the protection and administration of the shared secret [5].
For this reason, the use of individual keys per device constitutes an important practice. Reusing the same PSK across multiple nodes would expand the consequences of an eventual secret exposure and reduce the compartmentalization provided by the point-to-point architecture.

This difference again reflects distinct objectives: TLS has adequate mechanisms for formal and scalable identity management, while the ESPHome model is especially convenient when nodes belong to a locally managed, centrally administered infrastructure.

## 7.4. Resource Constraints and Scope of Results

The measurements analyzed confirm that the cost of cryptographic mechanisms depends significantly on the platform and its implementation.

On ESP32, for example, Espressif documentation shows that the approximate Mbed TLS heap usage can vary from 42,196 bytes with a default configuration to around 22,013 bytes when dynamic buffer allocation mechanisms are used [21].

Similarly, the improvements incorporated in ESPHome 2026.9 considerably modified the cost of implementing Noise on ESP8266 and ESP32 [20].

These results show that it is not appropriate to associate a fixed latency or memory value with TLS or Noise independently of the library, software version, processor, and configuration used.

For this reason, the values presented in this work should be interpreted as evidence obtained from published implementations and measurements, and not as a direct experimental comparison performed under identical conditions. This limitation follows directly from the methodology described in Section 3.6 and constitutes the principal constraint on the quantitative conclusions of the review.

A conclusive quantitative validation would require implementing both mechanisms on the same device and subjecting them to equal network conditions, payload sizes, transmission intervals, and reconnection patterns. This limitation is at the same time an opportunity to extend the study later through a controlled test bench.

### 7.5. Implications for an Educational Home Automation Infrastructure

In the context of a home automation infrastructure oriented toward classrooms at the Tecnológico Nacional de México, Campus Veracruz, the analysis allows establishing technical criteria for selecting the communication mechanism according to the function of each node.

Devices intended exclusively to interact with Home Assistant can benefit from the direct integration provided by the encrypted ESPHome Native API, particularly when using resource-constrained microcontrollers and seeking to reduce provisioning complexity.

When the information generated by the devices must be shared with other systems, stored by external platforms, used by analytics tools, or distributed among different applications, MQTT over TLS can provide a more flexible architecture.

A real infrastructure could, therefore, incorporate both mechanisms at different levels, selecting them according to the interoperability, security, administration, and computational capacity needs of each component.

## 8. CONCLUSIONS

The analysis performed allowed examining the security and robustness characteristics of two encrypted communication mechanisms used in local home automation architectures based on ESP-family microcontrollers and Home Assistant: MQTT protected through TLS 1.3 and the ESPHome Native API with Noise-based encryption. The review was conducted as a systematized review with explicit research questions, declared inclusion and exclusion criteria, and an identified threat model, as described in Section 3.

The results show that both mechanisms provide confidentiality and integrity for communications, but they respond to different architectural and trust management models. MQTT over TLS 1.3 relies on a publish-subscribe architecture mediated by a broker and can be integrated with formal authentication mechanisms, digital certificates, PKI infrastructure, and granular authorization controls [3], [4], [11]. These characteristics favor its use in environments where multiple producers and consumers participate or where interoperability between heterogeneous platforms is required. The encrypted ESPHome Native API uses a direct communication architecture with Home Assistant and a specific cryptographic suite based on the Noise_NNpsk0_25519_ChaChaPoly_SHA256 pattern [5], [12]. This scheme reduces the number of cryptographic decisions during configuration and uses a 32-byte pre-shared key as the central element of the trust model. Its direct integration and the optimizations introduced for ESP platforms make it particularly appropriate for nodes whose main purpose is to exchange states and commands with Home Assistant.

The analysis of the implementations also confirms that the computational cost of cryptographic mechanisms cannot be expressed through universal values. Memory consumption, session establishment time, and code footprint depend on factors such as the microcontroller used, the cryptographic library, the selected configuration, and the software version [9], [10], [20], [21]. The improvements recently incorporated into ESPHome demonstrate that changes in the implementation can significantly modify the cost of encryption without altering the underlying cryptographic protocol.

From a security perspective, it is also not appropriate to associate a vulnerability directly with an algorithm without considering its implementation [15]–[18]. The cases reviewed related to AES show that certain timing or side-channel vulnerabilities have been a consequence of specific software or hardware implementations and not of a cryptographic weakness inherent to the algorithm. This aspect is particularly relevant in embedded systems, where processor, memory, and cryptographic accelerator characteristics directly influence the practical resistance of the implementation.

It is important to reiterate the principal limitation of this review: the quantitative values presented for MQTT over TLS 1.3 and for the ESPHome Native API come from different hardware platforms, firmware versions, and library configurations, and the evidence for the ESPHome Native API relies substantially on official project documentation rather than on independent peer-reviewed studies. Consequently, the resource-consumption figures reported in this work should be interpreted as a consolidated reference of published measurements and not as an experimental comparison performed under identical conditions.

For a local home automation infrastructure oriented toward classrooms at the Tecnológico Nacional de México, Campus Veracruz, the results allow proposing a selection based on the operating scenario. Nodes intended exclusively to integrate with Home Assistant can use the encrypted ESPHome Native API as a direct communication mechanism, while MQTT over TLS 1.3 is appropriate when information must be distributed among multiple applications, services, or external platforms.

Consequently, the selection between the two mechanisms should not be formulated as a decision based on absolute superiority, but as an architectural choice conditioned by the interoperability, identity management, resource consumption, scalability, and trust model requirements of the system.

As future work, experimental validation on a common test bench is proposed to allow running both mechanisms under equivalent conditions. This evaluation should consider, at minimum, session establishment time, dynamic memory used, total size of transmitted frames, message processing time, and energy consumption. Applying the same set of tests on ESP8266 and different ESP32 variants would allow complementing the systematized review presented in this work with directly comparable experimental evidence and would address the principal limitation identified in Section 3.6.

## REFERENCES

[1] H. Fereidouni, O. Fadeitcheva, and M. Zalai, "IoT and man-in-the-middle attacks," Security and Privacy, vol. 8, no. 2, e70016, 2025. https://doi.org/10.1002/spy2.70016
[2] V. De Angelis, "Man-in-the-middle attacks against IoT devices," Tutorial, Traffic Measurement and Analysis Conference (TMA 2024), 2024. https://tma.ifip.org/2024/
[3] OASIS, "MQTT Version 5.0," OASIS Standard, 2019. https://docs.oasis-open.org/mqtt/mqtt/v5.0/mqtt-v5.0.htm
[4] S. U. A. Laghari, W. Li, S. Manickam, P. Nanda, A. K. Al-Ani, and S. Karuppayah, "Securing MQTT ecosystem: Exploring vulnerabilities, mitigations, and future trajectories," IEEE Access, 2024. https://doi.org/10.1109/ACCESS.2024.3412030
[5] ESPHome, "ESPHome API protocol details," Open Home Foundation, 2026. https://developers.esphome.io/architecture/api/protocol_details/

[6] M. J. Grant and A. Booth, "A typology of reviews: An analysis of 14 review types and associated methodologies," Health Information & Libraries Journal, vol. 26, no. 2, pp. 91–108, 2009.

[7] C. Okoli, "A guide to conducting a standalone systematic literature review," Communications of the Association for Information Systems, vol. 37, art. 43, 2015.

[8] V. Smith, D. Devane, M. Begley, and M. Clarke, "Methodology in conducting a systematic review of systematic reviews of healthcare interventions," BMC Medical Research Methodology, vol. 11, art. 15, 2011

[9] G. Restuccia, H. Tschofenig, and E. Baccelli, "Low-power IoT communication security: On the performance of DTLS and TLS 1.3," arXiv, 2020. https://doi.org/10.48550/arXiv.2011.12035

[10] T. Prantl, L. Iffländer, S. Herrnleben, S. Engel, S. Kounev, and C. Krupitzer, "Performance impact analysis of securing MQTT using TLS," in Proc. 2021 ACM/SPEC Int. Conf. Performance Engineering, 2021, pp. 241–248. https://doi.org/10.1145/3427921.3450253

[11] E. Rescorla, "The Transport Layer Security (TLS) Protocol Version 1.3," RFC 9846, Internet Engineering Task Force, 2026. https://doi.org/10.17487/RFC9846

[12] T. Perrin, "The Noise Protocol Framework (Revision 34)," 2018. https://noiseprotocol.org/noise.html

[13] [13] National Institute of Standards and Technology, "Advanced Encryption Standard (AES)," FIPS PUB 197, 2001.

[14] D. J. Bernstein, "ChaCha, a variant of Salsa20," in Workshop Record of SASC 2008: The State of the Art of Stream Ciphers, 2008.

[15] National Vulnerability Database, "CVE-2024-45191: Matrix libolm AES cache-timing vulnerability," National Institute of Standards and Technology, 2024. https://nvd.nist.gov/vuln/detail/CVE-2024-45191

[16] National Vulnerability Database, "CVE-2018-0498: Mbed TLS cache-based side-channel vulnerability," National Institute of Standards and Technology, 2018. https://nvd.nist.gov/vuln/detail/CVE-2018-0498

[17] Espressif Systems, "Security Advisory Concerning Breaking the Hardware AES Core and Firmware Encryption of ESP32 Chip Revision v3.0," AR2022-003, Version 2.0, 2022.

[18] Espressif Systems, "Security Advisory Concerning AES Key Recovery Using Voltage Fault Injection on ESP32-S3," AR2026-005, 2026.

[19] K. Kwiatkowski, P. Kampanakis, B. E. Westerbaan, and D. Stebila, "Post-Quantum Traditional (PQT) Hybrid Key Agreement Mechanisms for TLS 1.3," RFC 10024, Internet Engineering Task Force, 2026. https://doi.org/10.17487/RFC10024

[20] ESPHome, "ESPHome 2026.9.0: Faster builds and encrypted updates," Open Home Foundation, Sep. 16, 2026. https://esphome.io/blog/2026/09/16/esphome-2026-9/

[21] Espressif Systems, "Mbed TLS — ESP-IDF Programming Guide," 2026. https://docs.espressif.com/projects/esp-idf/en/latest/esp32/api-reference/protocols/mbedtls.html.